\documentclass[letterpaper]{article} % DO NOT CHANGE THIS
\usepackage{aaai2027}  % DO NOT CHANGE THIS
\usepackage[hyphens]{url}  % DO NOT CHANGE THIS
\usepackage{graphicx} % DO NOT CHANGE THIS
\usepackage{natbib}  % DO NOT CHANGE THIS AND DO NOT ADD ANY OPTIONS TO IT
\usepackage{caption} % DO NOT CHANGE THIS AND DO NOT ADD ANY OPTIONS TO IT
\usepackage{booktabs}
\usepackage{amsfonts}
\usepackage{amsmath}
\usepackage{array}
\usepackage{pifont}
\usepackage{subcaption}
\usepackage{colortbl} % xcolor is already loaded by aaai2027.sty

\definecolor{osshade}{RGB}{242,248,251}
\definecolor{osshadelt}{RGB}{251,251,251}
\newcommand{\ourrow}{\rowcolor{osshadelt}}
\newcolumntype{G}{>{\columncolor{osshadelt}}c}
\newcolumntype{Q}{>{\columncolor{osshade}}l}

\newcommand*\OK{\ding{51}}

\title{OmniSpectra: A Unified Foundation Model for Universal Spectra \\ Representation Learning}

\author{
    Md Khairul Islam\textsuperscript{\rm 1},
    Zeyu Xia\textsuperscript{\rm 1},
    Adam S. Bolton\textsuperscript{\rm 2},
    Judy Fox\textsuperscript{\rm 1,3}
}
\affiliations{
    \textsuperscript{\rm 1}Department of Computer Science, University of Virginia\\
    \textsuperscript{\rm 2}SLAC National Accelerator Laboratory\\
    \textsuperscript{\rm 3}School of Data Science, University of Virginia\\
    mi3se@virginia.edu, zeyu.xia@virginia.edu, abolton@slac.stanford.edu, ckw9mp@virginia.edu
}

\begin{document}

\maketitle

\begin{abstract}

Standard sequence models encode position by token rank rather than physical coordinate, typically forcing practitioners to destructively resample continuous physical signals such as spectra observed across heterogeneous instruments. We identify this coordinate mismatch as a structural limitation in cross-instrument representation learning, and propose OmniSpectra, a wavelength-native foundation model for spectral representation learning: a continuous-coordinate masked autoencoder that discards the integer pixel indexing of sequence modeling. Mapping physical coordinates to a logarithmic basis turns the multiplicative scaling induced by cosmological redshift into a uniform additive shift. In continuous rotary position embeddings, this shared shift cancels from relative token distances rather than producing grid-dependent positional distortion. We further counter the spectral bias inherent to implicit neural decoders with a physics-grounded, resolution-gated Fourier decoder that provides anti-aliasing. Pretrained on eight surveys and benchmarked against seven baselines, our model reaches state-of-the-art performance with only 12.2M parameters while outperforming substantially larger models, indicating that encoding physical structure can substitute for parameter scale.

% Standard sequence models encode position by token rank rather than physical coordinate, often requiring destructive resampling of continuous physical signals such as spectra from heterogeneous instruments. We identify this coordinate mismatch as a structural limitation in cross-instrument representation learning, and propose OmniSpectra, a wavelength-native foundation model: a continuous-coordinate masked autoencoder that discards integer pixel indexing. In logarithmic wavelength coordinates, cosmological redshift becomes an additive shift whose shared offset cancels from the relative positional differences used by continuous rotary embeddings, avoiding grid-dependent token-index distances. We further address the spectral bias of implicit neural decoders with a Fourier decoder whose basis is attenuated above each instrument's measured resolution limit. Pretrained on eight surveys and benchmarked against seven baselines, OmniSpectra reaches state-of-the-art performance with only 12.2M parameters while outperforming substantially larger models, indicating that encoding physical structure can substitute for parameter scale.

\end{abstract}

\section{Introduction}%
\label{sec:intro}

A central challenge in representation learning is building one model that generalizes across data sources with heterogeneous structure, resolution, coordinate systems, or sampling density, without dataset-specific preprocessing. Astronomy presents this at scale: spectroscopic surveys already cover hundreds of millions of objects across many instruments and wavelength regimes~\citep{audenaert2024multimodal}, and the coming decade will add petabytes of high-fidelity data. The heterogeneity forces a compromise: either per-instrument models, or homogenization by lossy resampling, which blurs narrow high-frequency features such as absorption and emission lines and injects interpolation artifacts into the measurements. Source classification, redshift estimation, and stellar and galaxy property prediction would all benefit from a shared representation, but only if it can ingest native spectra from heterogeneous instruments without resampling.

Standard sequence models learn structural relationships purely from data, which is sample-inefficient and leaves the representation tied to each instrument's sampling grid. This alignment challenge has driven a wave of multimodal astronomical foundation models: contrastive frameworks~\cite{parker2024astroclip}, omnimodal architectures with modality-specific tokenizers~\cite{parker2026aion}, and generative autoencoders~\cite{melchior2023autoencoding}.

\begin{figure*}[!t]
  \centering
  \begin{subfigure}{0.4\textwidth}
    \centering
    \includegraphics[width=\textwidth]{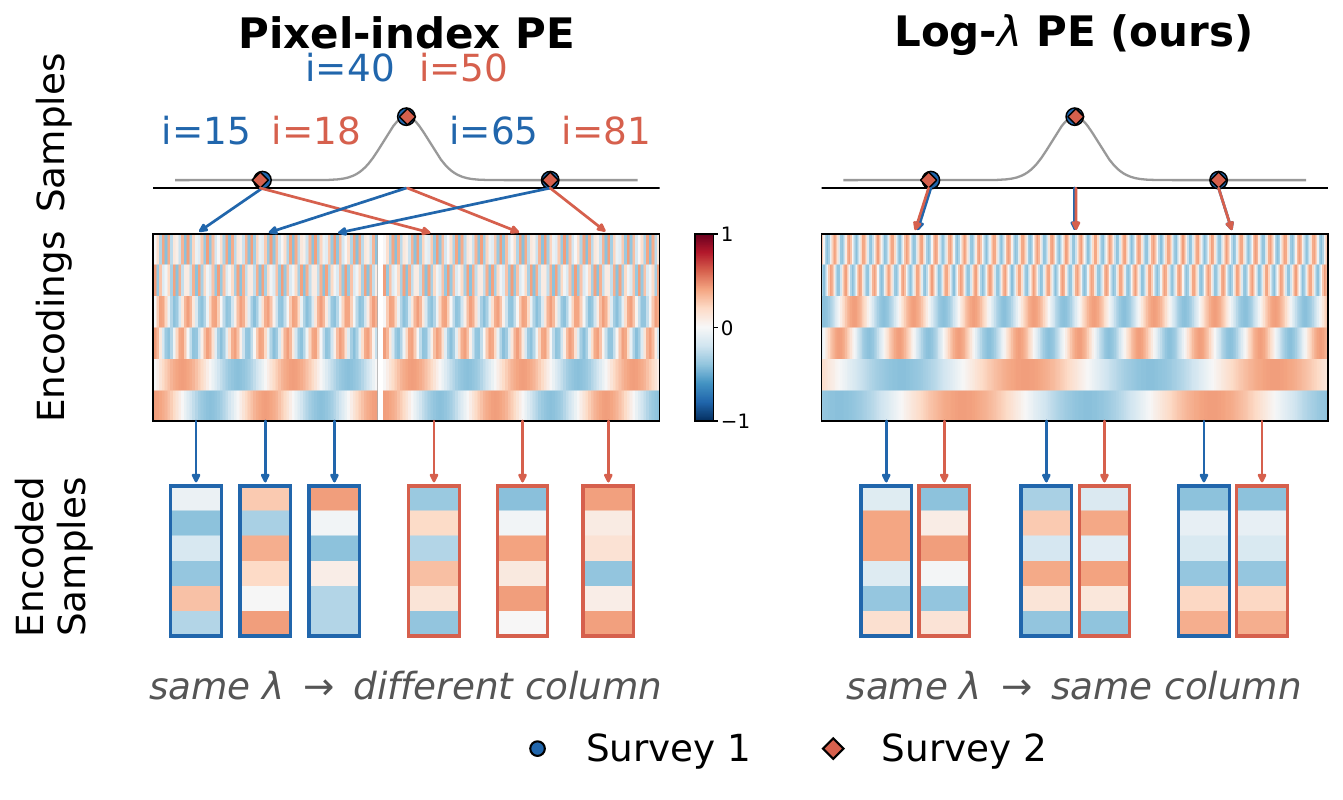}
    \subcaption{Positional encoding of H$\alpha$, two surveys.}
    \label{fig:pe_encoding}
  \end{subfigure}
  \hfill
  \begin{subfigure}{0.47\textwidth}
    \centering
    \includegraphics[width=\textwidth]{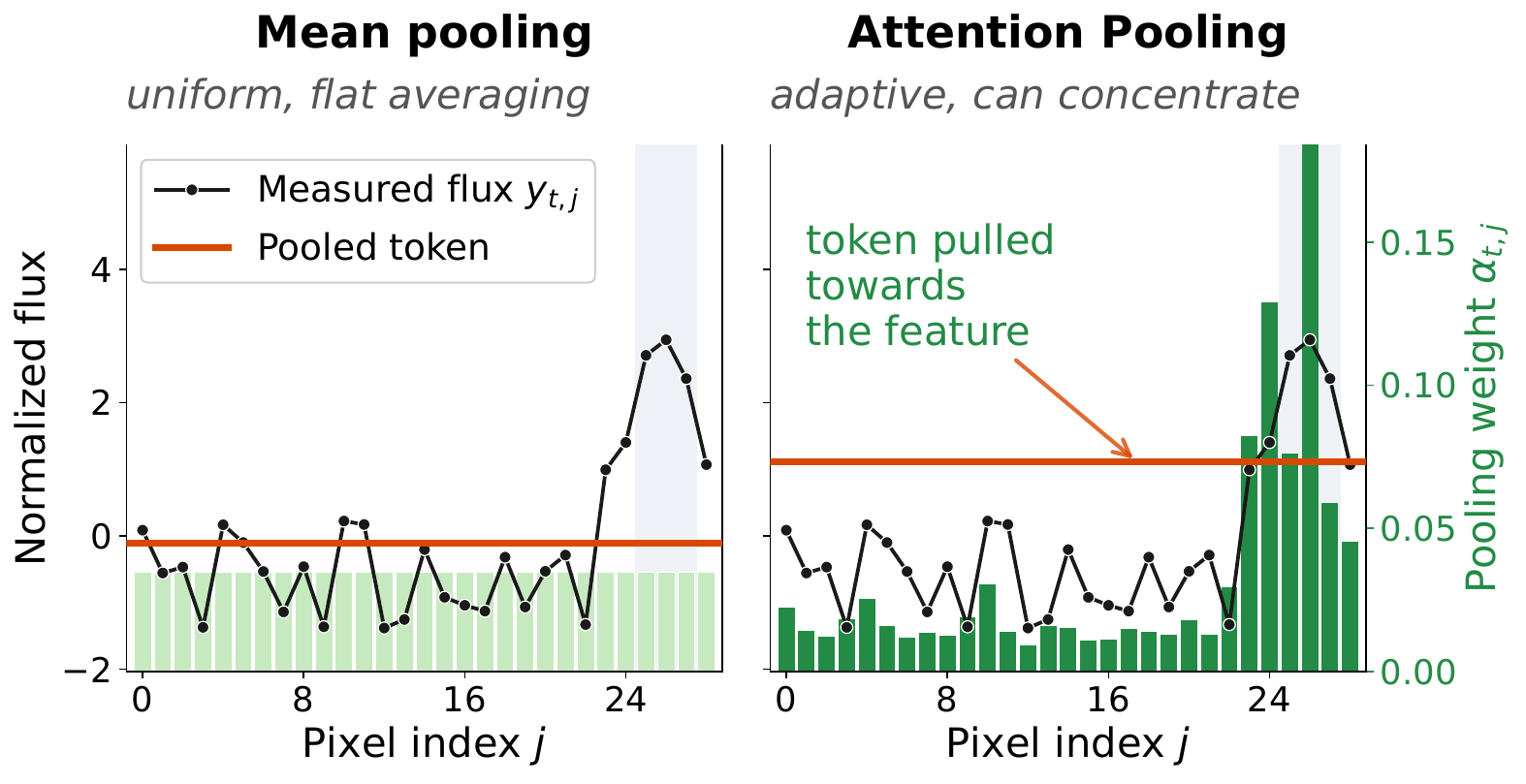}
    \subcaption{Attention pooling on a real DESI patch.}
    \label{fig:pooling_toy_real}
  \end{subfigure}
  \caption{(a) Pixel-index PE (\emph{left}) gives one wavelength a different code in each survey; $\log\lambda$ PE (\emph{right}) gives it the same code. (b) Mean pooling (\emph{left}) weights all $P$ pixels equally, diluting a narrow feature with the continuum; attention pooling (\emph{right}, Eq.~\ref{eq:attnpool}) concentrates the token on it.}
  \label{fig:tokenizer_pe_illustration}
\end{figure*}

A critical obstacle remains: instruments sample the same wavelength range at different pixel densities. When tokenization and positional encoding are defined over the pixel index, a feature such as the H$\alpha$ line falls at a different pixel on each detector (Fig.~\ref{fig:pe_encoding}), so one physical wavelength gets a different code per survey. The mismatch is structural: even when tokens carry continuous wavelength, attention geometry stays tied to sequence index, so cross-survey weight sharing requires lossy resampling and representations transfer poorly.

\begin{table}[!ht]
  \centering
  \small
  \setlength{\tabcolsep}{3pt}
  \begin{tabular}{@{}llQ@{}}
  \toprule
   \textbf{Dimension} & \textbf{Std Transformer} & \textbf{OmniSpectra} \\
  \midrule
  Content & Semantic tokens & \textbf{Local line shape} \\
  Position Anchor & Integer index $i$ & \textbf{Physical $\log\lambda$} with \textbf{RoPE}\\
  Mapping & Sequence steps & \textbf{Continuous $\log\lambda$ axis} \\
  Attention & Distance $(i-j)$ & \textbf{Energy ratio} $\log(\lambda_A/\lambda_B)$ \\
  Decoder bias & Low-frequency bias & \textbf{Resolution gating} \\
  \bottomrule
  \end{tabular}
  \caption{The core inductive biases: standard Transformer sequence modeling vs.\ OmniSpectra.}
  \label{tab:inductive_biases}
\end{table}

\begin{figure*}[!ht]
  \centering
  \includegraphics[width=0.8\linewidth]{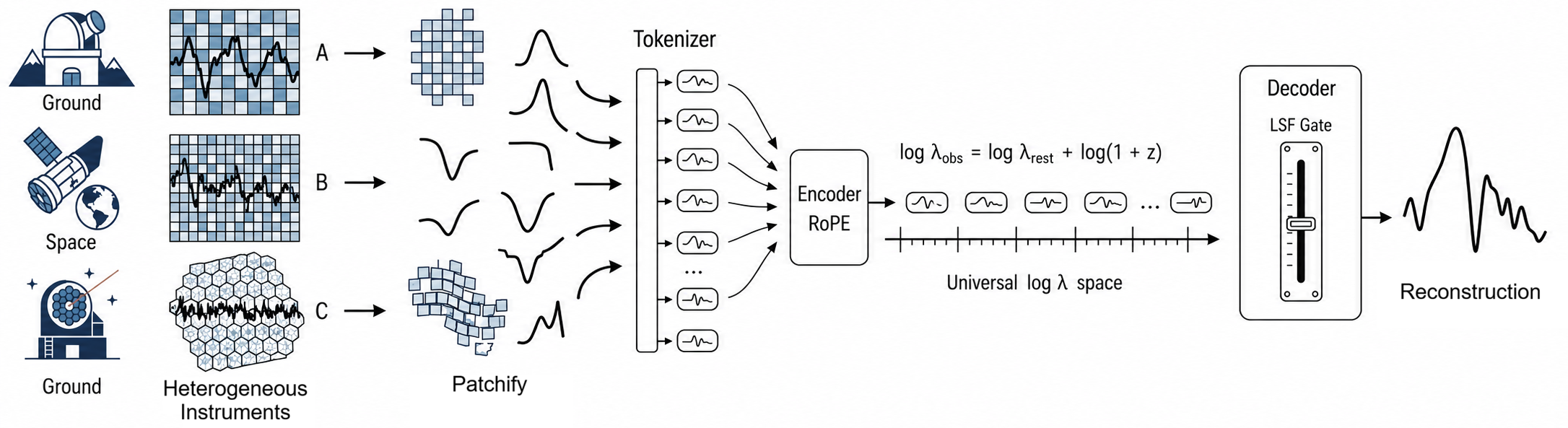}
  \caption{OmniSpectra architecture. The encoder extracts local morphology and places tokens in continuous physical log-wavelength ($\log\lambda$) space, where redshift dilation becomes a translation. The cross-attention Fourier decoder reconstructs at native coordinates, with an LSF gate tied to the instrument's resolution.}
  
  \label{fig:architecture}
\end{figure*}
\textbf{Our approach.} We introduce \textbf{OmniSpectra} (Fig.~\ref{fig:architecture}), a \emph{wavelength-native} approach to representation learning that replaces integer pixel indexing with physically anchored logarithmic wavelength coordinates. It factorizes the problem: hardware-efficient fixed-pixel spans extract local line shape from discrete detectors, while a shared continuous $\log \lambda$ space unifies the resulting tokens, so attention operates in physical rather than detector space. Coupled with a resolution-aware implicit Fourier decoder, one set of weights ingests heterogeneous spectra without a fixed grid.
OmniSpectra's core innovation is evaluating token distance in $\log\lambda$ space (Table~\ref{tab:inductive_biases}), reducing positional distance to the physical ratio of wavelengths and grounding a parameter-efficient, continuous-coordinate attention geometry in physics.
%Two physically motivated inductive biases---(1) \textbf{mapping cosmological redshift to an additive shift in $\log\lambda$} and (2) \textbf{bounding the decoder by the spectrograph's line-spread function to prevent aliasing}---offer a more parameter-efficient and robust alternative to scaling model size. 
Our contributions are threefold: (i) \textbf{a unified wavelength-anchored positional encoding}, realized with continuous 1D Rotary Position Encoding (RoPE), that unifies heterogeneous instruments without interpolation; (ii) \textbf{a resolution-gated implicit Fourier decoder} that reconstructs at each instrument's native coordinates with reduced discretization artifacts or hallucinations; and (iii) \textbf{a unified multi-instrument pretraining framework} that handles heterogeneous grids efficiently. Pretrained on 6.4M spectra from eight surveys, OmniSpectra matches or exceeds specialized models on \textit{source classification}, \textit{stellar} and \textit{galaxy property estimation}, and \textit{redshift prediction}, without resampling or per-instrument adaptation.

\section{Related Work}
\label{sec:background}

\textbf{Continuous-coordinate models.}
Perceiver~\citep{jaegle2021perceiver} and NaViT~\citep{dehghani2023patch} accept variable-length inputs but still identify observations by index-based positional codes, so one physical feature gets different codes across sampling densities. Relative-position methods such as RoPE~\citep{su2024roformer} and ALiBi~\citep{press2021train} improve length generalization by conditioning attention on distances between token indices, but token rank stays instrument-dependent, leaving cross-instrument misalignment unresolved. Neural operators~\citep{li2021fourier}, implicit neural representations~\citep{sitzmann2020implicit}, and set-based models~\citep{lee2019set,zaheer2017deepsets} instead act directly on continuous coordinates and accommodate irregular sampling. The same mismatch recurs across the sciences: instrument-agnostic Earth-observation models condition on each band's physical wavelength~\citep{xiong2024neural}, neural implicit methods ingest raw measurement coordinates in MRI~\citep{huang2023neural}, and mass-spectrometry models embed each peak's continuous $m/z$~\citep{bushuiev2026self}, while fixed-grid laboratory spectroscopy transfers poorly~\citep{koddenbrock2026ramanbench}. 

% None targets pretraining across heterogeneous spectroscopic instruments. OmniSpectra bridges these directions with a masked autoencoder whose encoder and decoder are defined entirely in terms of the physical wavelength coordinate $\lambda$.

\textbf{Physics-informed representations.}
Domain structure in learned representations can improve sample efficiency and out-of-distribution generalization~\citep{wang2023scientific,thomas2018tensor}; for spectra it must capture both the location and the fine-scale structure of the signal. Fourier feature networks~\citep{tancik2020fourier} address \emph{spectral bias}~\citep{rahaman2019spectral}: standard MLPs fit low frequencies first, a property of the network's own Fourier spectrum, not the observed spectrum. This lets coordinate-conditioned decoders recover narrow, high-frequency features. OmniSpectra uses this principle to reconstruct flux continuously at each instrument's native wavelength coordinates.

\textbf{Spectroscopic models.}
Most spectroscopic foundation models remain tied to one instrument and its fixed pixel grid. AstroCLIP~\citep{parker2024astroclip} encodes fixed-length spectra on the native detector grid, but its positional encoding carries no wavelength information; GalSpecNet~\citep{wu2024galaxy} and SpecPT~\citep{pattnaik2025specpt} likewise perform strongly within individual surveys without addressing cross-instrument sampling. Spender~\citep{melchior2023autoencoding} incorporates physical structure analytically, using the known redshift $z$ to resample observations onto a common rest-frame grid; this avoids learning the transformation, but requires $z$ at training and inference and stays tied to a fixed grid. Recent work addresses broader heterogeneity: AION-1~\citep{parker2026aion} tokenizes each instrument separately, and \citet{shen2025} pretrain one encoder--decoder on native grids with an absolute sinusoidal embedding of $\lambda$. OmniSpectra instead makes the coordinate relative and logarithmic, so redshift is a uniform translation in attention.

\section{Methodology}%
\label{sec:model}
% OmniSpectra is a masked autoencoder whose tokenization, positional encoding, and reconstruction are all defined on the physical coordinate $\lambda$ rather than on pixel or patch index (Fig.~\ref{fig:architecture2}). The architecture follows a single factorization principle: \emph{every token is a content--position pair}. Content is produced by a patch tokenizer that attention-pools native pixels, capturing local line morphology regardless of sampling density; position is a continuous encoding of the patch's physical wavelength, embedding the tokens of every instrument on one shared $\log\lambda$ axis. No quantity entering the encoder or decoder depends on the pixel index.

\subsection{Problem Statement}
\label{sec:formulation}
% \textbf{The Mathematics of Spectral Observations}.
A spectrum is a discrete observation of the continuous function $f:\mathbb{R}_{>0}\to\mathbb{R}$ at instrument-specific coordinates. Let $\mathcal{D}=\{\mathcal{S}^{(n)}\}_{n=1}^{N}$ be $N$ spectra from multiple instruments, each of which
\begin{equation}
\mathcal{S}^{(n)} = \bigl\{(\lambda_i,\, f_i,\, \sigma_i^{\mathrm{LSF}})\bigr\}_{i=1}^{L^{(n)}}
\label{eq:spectrum}
\end{equation}
consists of $L^{(n)}$ coordinate-value pairs, where $\lambda_i\in\mathbb{R}_{>0}$ is the physical coordinate, $f_i\in\mathbb{R}$ the signal value, and $\sigma_i^{\mathrm{LSF}}>0$ the instrument's \emph{line-spread function} (LSF) width at $\lambda_i$, i.e.\ the local resolution: how broad an infinitely narrow line appears after the spectrograph's optics, supplied as metadata rather than estimated. Both $L^{(n)}$ and the coordinate spacing $\Delta\lambda$ are fixed by the instrument rather than the signal; instruments in our corpus differ by up to $25\times$ in sampling density.

\subsection{Attention-Pooled Patch Tokenizer}
\label{sec:tokenizer}

A spectrum is divided into patches of $P$ consecutive pixels, reindexing its $L^{(n)}$ pixels by patch $t\in\{1,\ldots,T^{(n)}\}$ and within-patch position $j\in\{1,\ldots,P\}$ in place of the flat index $i$ of Eq.~\ref{eq:spectrum}; a validity mask $\mathbf{m}_t\in\{0,1\}^P$ excludes invalid pixels. Because $P$ is fixed in pixels rather than in $\lambda$, $T^{(n)}$ varies with instrument resolution: one patch spans a wide wavelength interval on a coarse grid but a narrow one on a fine grid. This stage is efficient but not grid-invariant; cross-instrument consistency comes downstream, from anchoring the positional code (Eq.~\ref{eq:gamma}) and reconstruction query (Eq.~\ref{eq:pq}) to $\lambda$.

Within each patch, the tokenizer extracts \emph{content}, the local feature shape, stripped of position, so the two recombine later on the physical axis. Each pixel is linearly embedded, $\mathbf{e}_{t,j}=\mathrm{Linear}(f_{t,j})\in\mathbb{R}^d$ with $d$ the embedding dimension, and a depthwise convolution mixes neighboring pixels to capture line widths and asymmetries: $\tilde{\mathbf{e}}_{t,j} = \mathrm{DWConv}(\mathbf{e}_t)_j$. A learned global query $\mathbf{q}\in\mathbb{R}^d$, shared across all patches and instruments, attention-pools these features into one token:

\begin{equation}
\alpha_{t,j} = \frac{m_{t,j}\exp\!\bigl(\mathbf{q}^\top \tilde{\mathbf{e}}_{t,j}/\sqrt{d}\bigr)}{\sum_{j'=1}^{P} m_{t,j'}\exp\!\bigl(\mathbf{q}^\top \tilde{\mathbf{e}}_{t,j'}/\sqrt{d}\bigr)},
\;\;
\mathbf{x}_t = \sum_{j=1}^{P} \alpha_{t,j}\,\tilde{\mathbf{e}}_{t,j}
\label{eq:attnpool}
\end{equation}

Mean pooling ($\alpha_{t,j}=1/P$) is the uniform-weight special case, which dilutes a narrow feature with the surrounding continuum; learned weights concentrate the token on it instead (Fig.~\ref{fig:pooling_toy_real}), and since absorption lines are flux minima, this selectivity is not merely a preference for large values. The resulting $\mathbf{x}_t$ carries no patch rank or instrument identifier.

% \begin{figure}[!ht]
%   \centering
%   \includegraphics[width=0.9\linewidth]{figures/pooling_toy_real_desi_gated.pdf}
%   \caption{Attention pooling on a real DESI patch. Mean pooling (\emph{left}) applies uniform weights $1/P$; attention pooling (\emph{right}, Eq.~\ref{eq:attnpool}) concentrates on locally informative pixels.}
%   \label{fig:pooling_toy_real}
% \end{figure}

\subsection{Log-Wavelength Positional Encoding}
\label{sec:pe}

The positional code supplies the other half of the factorization: where each token lies in physical wavelength. Each patch is anchored at its masked mean coordinate, and its code is a continuous function of it:
\begin{equation}
\mathbf{p}_t = \gamma\!\bigl(\log\bar\lambda_t - \log\lambda_0;\;\boldsymbol\omega\bigr)
\label{eq:pe}
\end{equation}
where $\lambda_0$ is a fixed global anchor shared across instruments, $\bar\lambda_t = \sum_j \lambda_{t,j}\,m_{t,j}\,/\,\lVert\mathbf{m}_t\rVert_1$ is the masked mean coordinate, and the encoding function $\gamma$ applies the interleaved sin/cos grouping \citet{vaswani2017attention} to $\log\lambda$ in place of token index:
\begin{equation}
\gamma(x;\boldsymbol\omega) = \bigl[\sin(\omega_k x),\;\cos(\omega_k x)\bigr]_{k=1}^{d/2}
\label{eq:gamma}
\end{equation}

% \begin{figure}[!ht]
%   \centering
%   \includegraphics[width=\columnwidth]{figures/pe_encoding_comparison.pdf}
%   \caption{Positional encoding (PE) of H$\alpha$ for two surveys of differing pixel density. Pixel-index PE \emph{(left)} maps the same wavelength to different codes; log$\lambda$ PE \emph{(right)} maps it to the same positional tensor on any instrument.}
%   \label{fig:pe_encoding}
% \end{figure}

The frequencies $\boldsymbol\omega$ are log-spaced over periods from sub-pixel to full-survey scales a continuous parameterization in the spirit of \citet{tancik2020fourier} rather than the transformer's fixed geometric schedule, so the code resolves both broad continuum structure and individual narrow lines. Unlike an index table, it can be evaluated at any wavelength (Fig.~\ref{fig:pe_encoding}).
\begin{figure*}[!ht]
  \centering
    \centering
    \includegraphics[width=\linewidth]{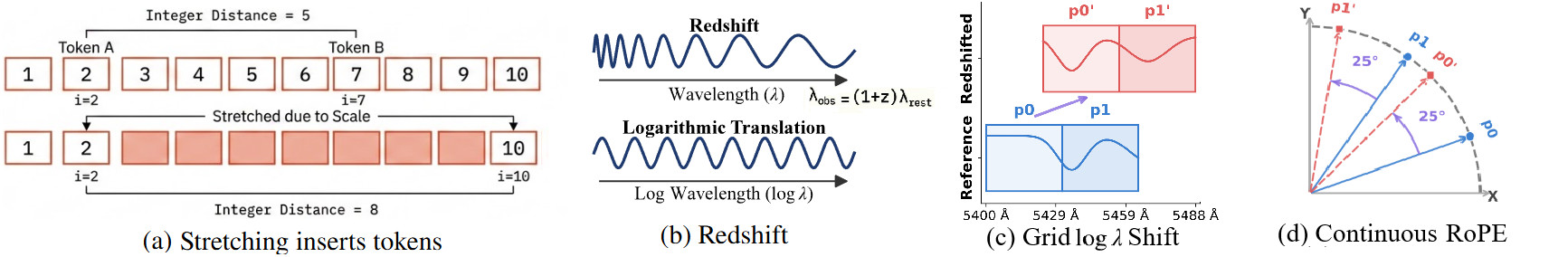}
  \caption{Redshift changes index distance \emph{(a)}, but adds the same offset to every $\log\lambda$ coordinate \emph{(b)}, shifting a windowed line profile uniformly on the log grid \emph{(c)}; continuous RoPE preserves the relative angle between tokens under this shift \emph{(d)}.}
  \label{fig:redshift_challenge}
\end{figure*}

\textbf{Tokens are embedded by wavelength, not sequence rank.}
In a standard Transformer, position is a lookup keyed by integer rank. Equation~\ref{eq:pe} instead maps wavelength deterministically onto one $\log\lambda$ axis: patches covering the same region receive comparable codes on any grid, up to the small residual difference in $\bar\lambda_t$ from pixel density. A denser grid places proportionally more tokens along that same axis rather than remapping them into an unrelated code space. OmniSpectra adds $\mathbf{p}_t$ to the content token, preserving absolute observer-frame wavelength.

%\begin{figure}[!ht]
%  \centering
%  \begin{subfigure}[b]{0.59\columnwidth}
%    \centering
%    \includegraphics[width=\linewidth]{figures/insert_tokens_diagram.png}
%    \subcaption{Stretching inserts tokens}
%    \label{fig:NLP}
%  \end{subfigure}
%  \hfill
%  \begin{subfigure}[b]{0.38\columnwidth}
%    \centering
%    \includegraphics{figures/unused/architecture/Figure 3 redshift.png}
%    \subcaption{Redshift}
%    \label{fig:redshift}
%  \end{subfigure}
%  \caption{A multiplicative wavelength stretch changes index distance \emph{(a)}, but becomes an additive displacement in $\log\lambda$ space \emph{(b)}.}
%  \label{fig:redshift_challenge}
%\end{figure}

Cosmological redshift maps every rest-frame wavelength according to $\lambda_{\mathrm{obs}}=(1+z)\,\lambda_{\mathrm{rest}}$, applying the same factor across the spectrum. Index position instead measures separation in token steps, so this mapping changes the number of intervening tokens and hence their positional distance by an amount set by the instrument's sampling grid (Fig.~\ref{fig:redshift_challenge}ab).

\textbf{Redshift is additive in log-wavelength.} In $\log\lambda$ the same mapping is additive,

\begin{equation}
\log\lambda_{\mathrm{obs}} = \log\lambda_{\mathrm{rest}} + \log(1+z)
\label{eq:redshift}
\end{equation}

so one redshift displaces every line by the same offset, the property that motivates uniform $\log\lambda$ binning in spectroscopic pipelines~\citep{bolton2012spectral} (Fig.~\ref{fig:redshift_challenge}c). We therefore also apply continuous 1D Rotary Position Embeddings (RoPE)~\citep{su2024roformer} over $\log\lambda$, which rotate queries and keys by an angle proportional to position, giving self-attention a relative-distance bias alongside the absolute code of Eq.~\ref{eq:pe} (Fig.~\ref{fig:redshift_challenge}d). Because the RoPE rotation depends only on relative distance, the shared offset cancels there for any pair of lines, e.g.\ $H\alpha$ and $H\beta$: $\bigl(\log\lambda_{H\alpha}+\log(1{+}z)\bigr)-\bigl(\log\lambda_{H\beta}+\log(1{+}z)\bigr)=\log(\lambda_{H\alpha}/\lambda_{H\beta})$.

% where $\lambda_{\mathrm{obs}}$ is observed wavelength, $z$ is the redshift factor, and $\lambda_{\mathrm{rest}}$ is rest wavelength. The rest-to-observed-frame mapping reduces to a linear translation: because the positional encoding is natively defined in log space, redshift becomes a uniform additive offset (Fig.~\ref{fig:redshift}) within the network's internal geometry.
% We therefore additionally apply continuous 1D Rotary Position Embeddings (RoPE)~\citep{su2024roformer} over the $\log\lambda$ coordinate, which rotates the query and key vectors by an angle proportional to position. 

% As the attention score depends only on the relative positional distance 
% %$ \Delta_{H\alpha,H\beta} = \log\lambda_{H\alpha}-\log\lambda_{H\beta}$ 
% (Eq.~\ref{eq:redshift}), redshift adds the same $\log(1+z)$ to both wavelengths and cancels: the relative positional distance $\Delta \log \lambda$ is invariant to redshift,

% Positional distance thus reduces to the physical ratio of wavelengths, $\log\lambda_A-\log\lambda_B = \log(E_B/E_A)$: relative attention distances are redshift-invariant and aligned with the fixed energy ratios of atomic transitions ($E = hc/\lambda$). The two codes are complementary: RoPE supplies the redshift-invariant relative prior, while Eq.~\ref{eq:pe} retains the absolute observer-frame wavelength that redshift estimation requires.

Positional distance thus reduces to the physical ratio of wavelengths, $\log\lambda_A-\log\lambda_B = \log(E_B/E_A)$, aligned with the fixed energy ratios of atomic transitions ($E = hc/\lambda$). The two codes divide it the way relative and absolute encodings divide position in sequence models generally: RoPE's relative term cancels it, so identifying a feature does not require already knowing $z$, while the absolute code of Eq.~\ref{eq:pe} retains it, since $z$ is itself a prediction target.

\begin{figure*}[!t]
  \centering
  \begin{subfigure}{0.38\textwidth}
    \centering
    \includegraphics[width=\textwidth]{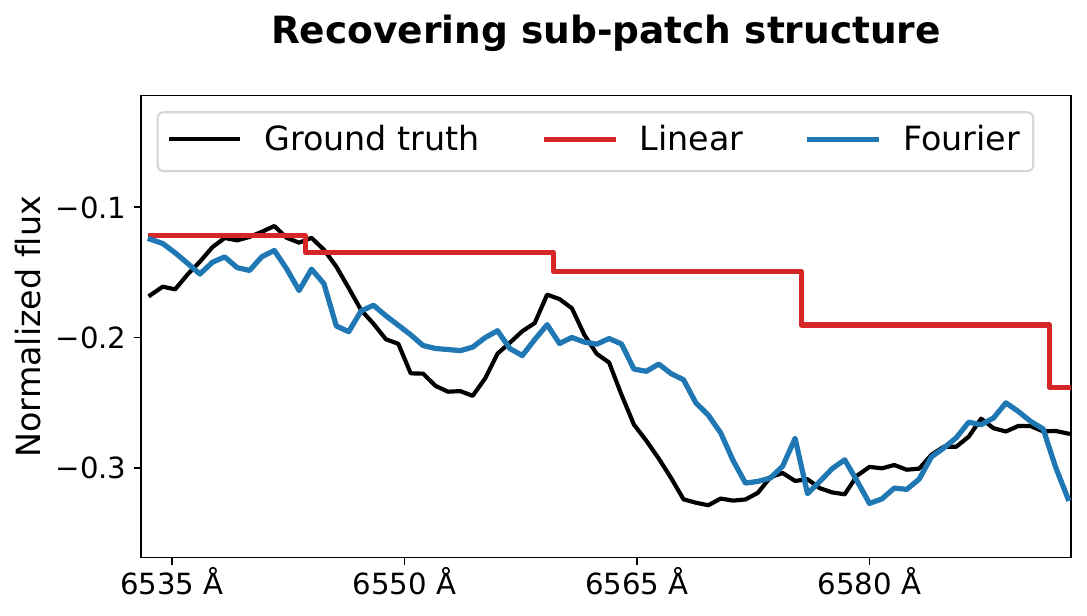}
    \subcaption{Fourier vs.\ linear decoding.}
    \label{fig:implicit_decoder}
  \end{subfigure}
  \hfill
  \begin{subfigure}{0.52\textwidth}
    \centering
    \includegraphics[width=\textwidth]{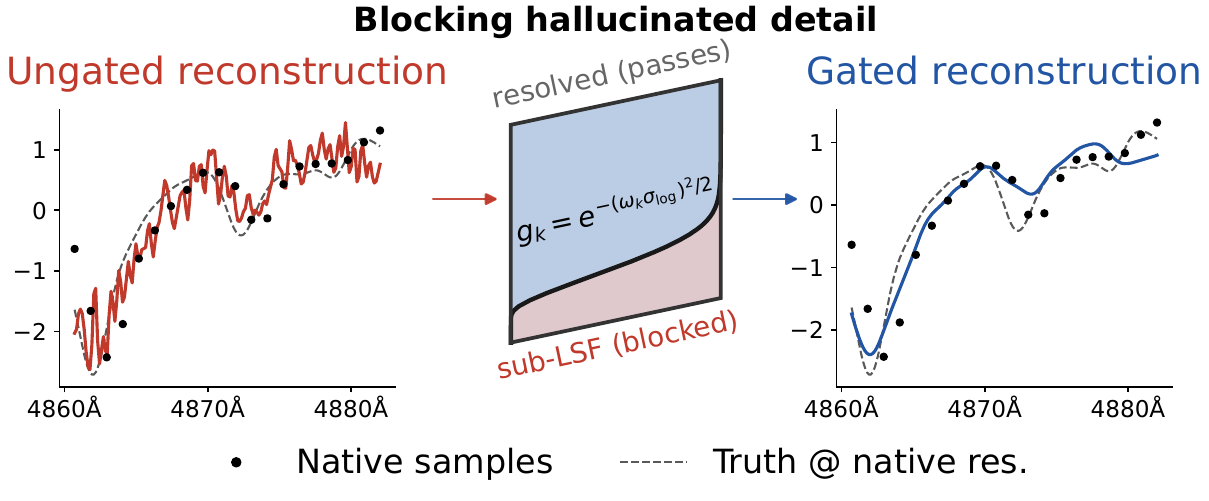}
    \subcaption{LSF gating, a SEGUE patch.}
    \label{fig:lsf_gate_mirror}
  \end{subfigure}
  \caption{The implicit decoder. (a) Linear and Fourier decoding from one shared encoder context. (b) The same model's ungated \emph{(left half)} and gated \emph{(right half)} reconstruction.}
  \label{fig:decoder_illustration}
\end{figure*}

\subsection{Implicit Resolution-Gated Fourier Decoder}
\label{sec:decoder}
Standard decoders predict one value per patch grid cell, producing discretization artifacts on narrow features and misalignment across differing pixel spacings. We instead query reconstruction at every native coordinate in two stages, for masked patch $q$ with centroid $\bar\lambda_q$ and pixel coordinates $\{\lambda_{q,i}\}$:

\textbf{Cross-attention over encoder tokens.} The query for patch $q$ cross-attends over the encoder tokens rather than a single pooled context, so reconstruction at $\bar\lambda_q$ draws on whichever tokens are physically relevant \citep{lee2023locality}. With $\mathbf{Z}\in\mathbb{R}^{T\times d}$ the encoder output and $\tilde{\mathbf{Z}}=\mathbf{Z}\mathbf{W}_{\!d}\in\mathbb{R}^{T\times d'}$ its projection to the decoder width $d'$, the query reuses Eq.~\ref{eq:gamma} at $\bar\lambda_q$, with $\boldsymbol\omega$ resampled to $d'/2$ values,

\begin{equation}
\mathbf{p}_q = \gamma(\log\bar\lambda_q - \log\lambda_0;\,\boldsymbol\omega)
\label{eq:pq}
\end{equation}

which then cross-attends to the projected sequence $\tilde{\mathbf{Z}}$:

\begin{equation}
\mathbf{h}_q = \mathrm{CrossAttn}(\mathbf{p}_q,\,\tilde{\mathbf{Z}})
\label{eq:crossattn}
\end{equation}
where $\mathbf{h}_q$ is the aggregated encoder context for patch $q$.

\textbf{Per-coordinate implicit decoding.} The context $\mathbf{h}_q$ combines with a resolution-gated Fourier basis evaluated at each native coordinate. For pixel $i$ of patch $q$, with $K_f$ log-spaced frequencies $\omega_k$ over the same range:
\begin{equation}
\boldsymbol\phi_{q,i} = \Bigl[g_k\cos(\omega_k\log\lambda_{q,i}),\;
                              g_k\sin(\omega_k\log\lambda_{q,i})\Bigr]_{k=1}^{K_f}
\label{eq:fourier}
\end{equation}

where the gate $g_k = \exp\!\bigl(-\tfrac{1}{2}(\omega_k\sigma_q^{\mathrm{LSF}})^2\bigr)$  attenuates frequencies above the effective resolution $\sigma_q^{\mathrm{LSF}}$ of patch $q$ (the mean per-pixel LSF width in log-$\lambda$ units), read from the instrument's reported resolution rather than tuned, following the Gaussian LSF model of fiber-spectrograph pipelines~\citep{bolton2010spectro}. One measured quantity thus achieves the anti-aliasing that Integrated Positional Encoding~\citep{barron2021mip} obtains by scaling a Fourier basis to the camera footprint, and that \citet{lee2023locality} approximate by learning bandwidths. Per-pixel coordinates enable sub-patch variation a linear decoder cannot express (Fig.~\ref{fig:implicit_decoder}); the prediction is:

\begin{equation}
\hat{f}(\lambda_{q,i}) = \mathrm{MLP}\!\bigl([\mathbf{h}_q;\;\boldsymbol\phi_{q,i}]\bigr)
\label{eq:mlp}
\end{equation}

% \begin{figure}[!ht]
%   \centering
%   \includegraphics[width=.8\linewidth]{figures/implicit_decoder_model.pdf}
%   \caption{Fourier vs.\ linear decoding from a shared encoder context. The linear decoder emits one value per patch (step function); the Fourier decoder tracks sub-patch variation at native coordinates.}
%   \label{fig:implicit_decoder}
% \end{figure}

\textbf{LSF gating discourages unresolved detail.} Attenuating unsupported frequencies in the basis itself, rather than leaving the network to learn to ignore them, limits the decoder's access to detail no measurement could produce. Fig.~\ref{fig:lsf_gate_mirror} passes one crossmatched SEGUE patch through the ungated and gated heads of a single trained model at $8\times$ native sampling: the ungated reconstruction develops between-pixel oscillations absent from an independent higher-resolution GALAH observation of the same star, which the gate substantially reduces. Because the final prediction uses a nonlinear MLP, this is an inductive bias, not a strict band-limit guarantee.

% \begin{figure}[!ht]
%   \centering
%   \includegraphics[width=0.9\linewidth]{figures/lsf/lsf_gate_mirror_samemodel.pdf}
%   \caption{The LSF gate blocks sub-resolution Fourier content. Left: ungated reconstruction of a crossmatched SEGUE patch shows fabricated high-frequency ripple; right: the same model's gated reconstruction.}
%   \label{fig:lsf_gate_mirror}
% \end{figure}

\section{Experimental Setup}
\label{sec:setup}

\textbf{Datasets.} We pretrain on 6.4M spectra from eight surveys: SDSS Legacy, BOSS, eBOSS, SEGUE-1, SEGUE-2, and DESI (optical), plus APOGEE and GALAH (near-infrared, high-resolution)~\citep{york2000sloan,eisenstein2011sdss,blanton2017sloan,abdurro2022seventeenth,buder2021galah,adame2024early}. These span resolving power $R\!\approx\!1{,}500$--$34{,}000$ (higher $R$ resolves finer detail) and 3,865--16,384 native pixels per spectrum. Every spectrum enters at its native grid, with no resampling or interpolation; per-survey statistics are in Table~A1.

% \textbf{Training.} OmniSpectra (12.2M parameters) is pretrained with a 0.5 mask ratio over $P{=}32$-pixel patches on 4 NVIDIA A100 GPUs; Table~A2 reports its full architecture and optimization settings.

% \textbf{Baselines.} We compare four pretrained models (AstroCLIP, AION-1, Shen et al., Perceiver) and two supervised single-instrument models (GalSpecNet, Spender), adding GaSNet-III~\citep{zhong2025galaxy} for redshift. We use public checkpoints where they exist for the evaluation instrument and train from scratch otherwise. Table~A2 gives their layer-level configurations and Table~A3 contrasts parameter counts, tokenization, positional encoding, resampling, and padding handling, the last of which matters because spectra of differing native length share a batch.
% Appendix letters shifted after the appendix was trimmed (its former Appendix C, "Formal Setup and Notation", is commented out there).
% \textbf{Training and baselines.} Training and baseline details are given in Appendices~D and~E of the technical appendix.
\textbf{Training and Baselines.} OmniSpectra has a 384-dim, 6-layer encoder and a 128-dim, 4-layer decoder (12.2M parameters). Each valid $P{=}32$-pixel patch is masked independently with probability 0.5. It is pretrained for 200 epochs on 4 NVIDIA A100 GPUs with AdamW (lr $10^{-4}$, weight decay 0.05, cosine annealing, batch size 64). We compare four pretrained models---AstroCLIP, AION-1, Shen et al., and Perceiver---two supervised single-instrument models, GalSpecNet and Spender, and GaSNet-III, a template-fitting $z$ estimator with no transferable representation, on redshift only. We pretrain Shen et al. and Perceiver ourselves on the identical corpus and collator under a matched step budget, so these are architecture-only comparisons. AION-1 and AstroCLIP are evaluated using public checkpoints, when available, on the same corpus. Full hyperparameters and baseline configurations are in Technical Appendices~C and~D.
\textbf{Transfer learning protocol.}
Foundation models are evaluated with a frozen encoder and a lightweight cross-attention pooling head that maps the token sequence to a fixed vector (following AION-1); supervised baselines (GalSpecNet, Spender) are trained end-to-end per task. We train 10 epochs with AdamW (lr=$10^{-4}$, weight decay 0.05), batch size 64, selecting the best model by validation loss.

\textbf{Downstream data splits.}
Galaxy property and redshift estimation use DESI EDR~\citep{adame2024early} cross-matched with DESI PROVABGS~\citep{hahn2023provabgs} (232.7K samples, 80:20 split); all other tasks use 12,500 random samples per instrument, 20\% held out. Results average over 5 random seeds.
%, with per-seed variance in Fig.~\ref{fig:low_data_regime}.

\textbf{Evaluation metrics.}
Source classification reports F1-score (robust to class imbalance): per-class in Table~\ref{tab:src_class_f1}, macro-averaged in the ablation (Table~\ref{tab:ablation}). Regression tasks report $R^2$ ($\uparrow$) and $\sigma^\mathrm{MAD} = 1.4826\,\mathrm{median}(|r - \mathrm{median}(r)|)$ ($\downarrow$), a robust dispersion estimate~\citep{leung2019deep}, with $r = \hat{y} - y$, or $r = (\hat{z}-z)/(1+z)$ for redshift. Redshift additionally reports the catastrophic-outlier rate, the percentage with $|r| > 0.15$~\citep{ilbert2009cosmos}, which $\sigma^\mathrm{MAD}$ by construction ignores~\citep{bolton2012spectral}.

\section{Results and Analysis}
\label{sec:results}

\subsection{Cross-Instrument Reconstruction}
\begin{figure}[!ht]
    \centering    \includegraphics[width=.88\columnwidth]{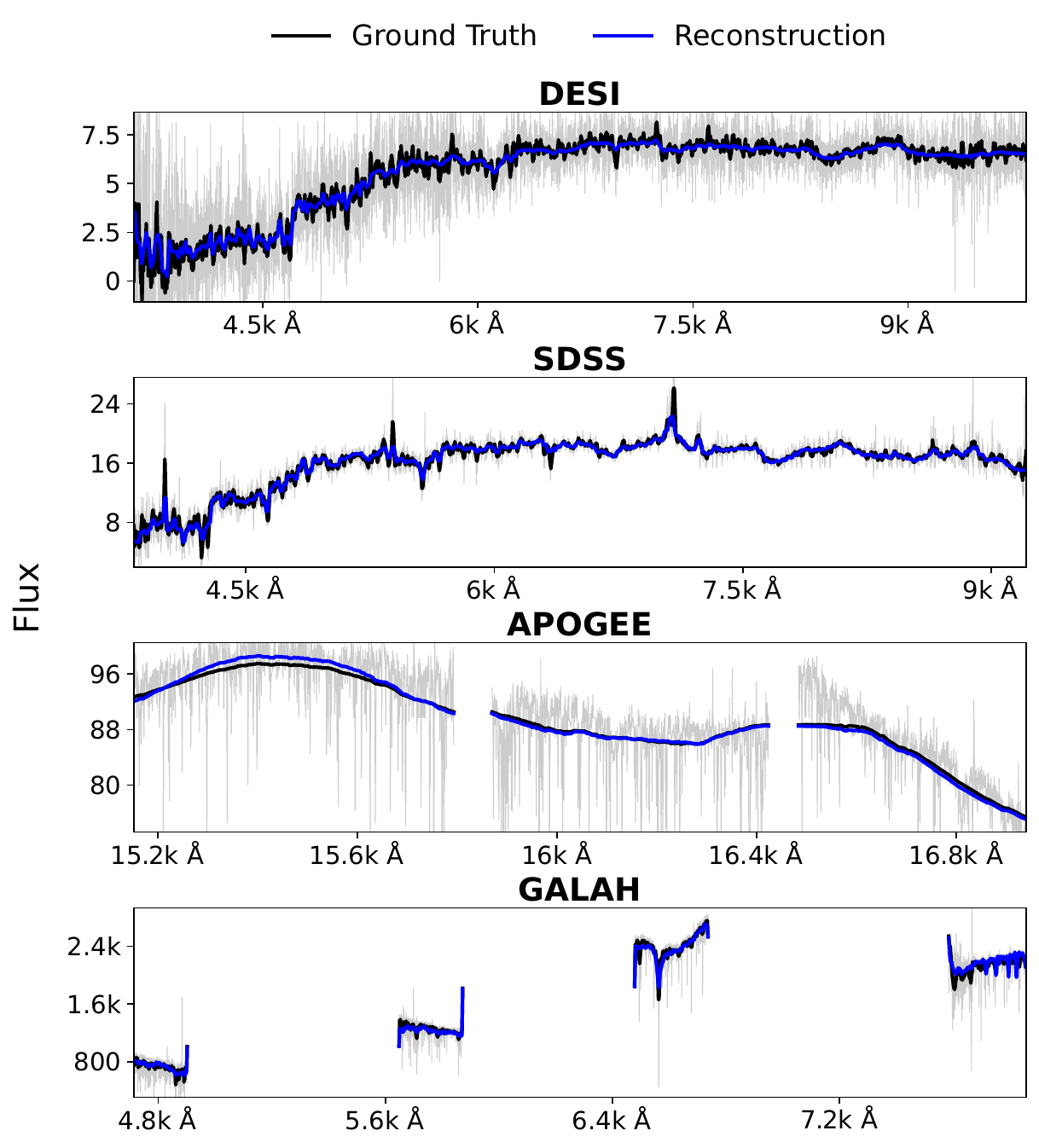}
    \caption{OmniSpectra's four-survey reconstructions with a single set of shared weights.}
    \label{fig:multisurvey_recon}
\end{figure}

Reconstruction is the precondition for every downstream result. Can shared weights return each instrument's measurements on its own grid? Fig.~\ref{fig:multisurvey_recon} decodes each survey at its own native wavelengths, and the reconstruction closely tracks the ground truth throughout

\begin{figure*}[!ht]
  \centering
  \includegraphics[width=\textwidth]{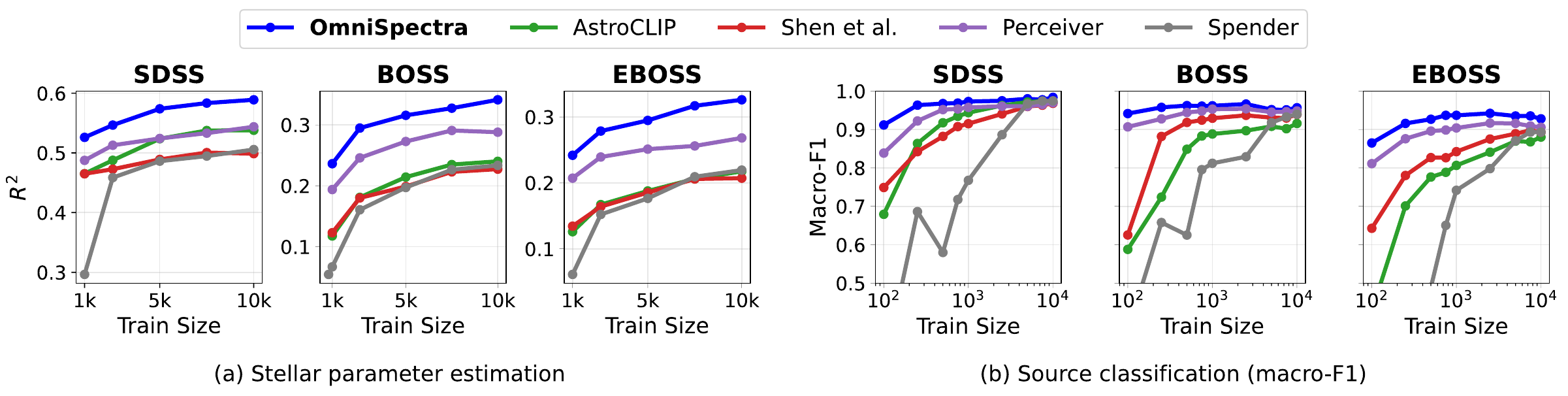}
  \caption{Low-data regime: average downstream performance.}
  \label{fig:low_data_regime}
\end{figure*}

\subsection{Stellar Parameter Estimation}

Stellar parameters test whether localized structure survives compression into frozen tokens, and whether one encoder serves instruments with disjoint coverage: three targets ($T_{\mathrm{eff}}$, $\log g$, $[\mathrm{Fe/H}]$) on SDSS \citep{york2000sloan} and APOGEE \citep{abdurro2022seventeenth}. OmniSpectra is best on all three for both instruments (Table~\ref{tab:star_props_by_target}) and remains strongest as labels are reduced (Fig.~\ref{fig:low_data_regime}(a)). It consistently keeps a large gap from the baselines as more labels are added. AION-1 is omitted here: its public spectral weights do not cover APOGEE, and on SDSS its official checkpoint trails every other pretrained model by a large margin.

\begin{table}[!ht]
  \centering
  \small
  \setlength{\tabcolsep}{4pt}
  \begin{tabular*}{\columnwidth}{l@{\extracolsep{\fill}}cccccc}
  \toprule
  & \multicolumn{2}{c}{$T_{\mathrm{eff}}$ (K)} & \multicolumn{2}{c}{$\log g$ (dex)} & \multicolumn{2}{c}{$[\mathrm{Fe/H}]$ (dex)} \\
  \cmidrule(lr){2-3}\cmidrule(lr){4-5}\cmidrule(lr){6-7}
  \textbf{Model} & $R^2$ & $\sigma^\mathrm{MAD}$ & $R^2$ & $\sigma^\mathrm{MAD}$ & $R^2$ & $\sigma^\mathrm{MAD}$ \\
  \midrule
  \multicolumn{7}{c}{\textit{SDSS}}\\
  \midrule
 \ourrow \textbf{OmniSpectra} & \textbf{0.93} & \textbf{158} & \textbf{0.31} & 0.22 & \textbf{0.58} & \textbf{0.21} \\
 %AION-1 & 0.12 & 1305 & 0.03 & 0.61 & 0.05 & 0.37 \\
  AstroCLIP & 0.90 & \underline{168} & 0.24 & 0.22 & 0.46 & \underline{0.21} \\
  Shen et al. & 0.89 & 200 & 0.19 & \underline{0.21} & 0.41 & 0.26 \\
  Perceiver & \underline{0.90} & 233 & \underline{0.22} & 0.31 & \underline{0.49} & 0.24 \\
  \midrule
  GalSpecNet & 0.72 & 238 & 0.13 & 0.21 & 0.33 & 0.29 \\
  Spender & 0.90 & 232 & 0.19 & \textbf{0.19} & 0.43 & 0.25 \\
  \midrule
  \multicolumn{7}{c}{\textit{APOGEE}}\\
  \midrule
  \ourrow \textbf{OmniSpectra} & \textbf{0.96} & \textbf{85} & \textbf{0.96} & \textbf{0.18} & \textbf{0.91} & \textbf{0.07} \\
  AstroCLIP & 0.94 & 100 & 0.92 & 0.25 & 0.83 & 0.10 \\
  Shen et al. & \underline{0.95} & \underline{92} & \underline{0.95} & \textbf{0.18} & \underline{0.87} & \underline{0.09} \\
  Perceiver & 0.82 & 217 & 0.76 & 0.44 & 0.55 & 0.20 \\
  \midrule
  GalSpecNet & 0.90 & 137 & 0.94 & \underline{0.20} & 0.83 & 0.09 \\
  Spender & 0.91 & 127 & 0.94 & 0.23 & 0.82 & 0.09 \\
  \bottomrule
  \end{tabular*}
 \caption{Stellar parameter estimation on SDSS (optical) and APOGEE (near-infrared). Ours is \colorbox{osshadelt}{shaded}; best/second best per block in \textbf{bold}/\underline{underline} (all tables).}
  % ; best/second best per block in \textbf{bold}/\underline{underline} (all tables)
  \label{tab:star_props_by_target}
\end{table}

\subsection{Galaxy Property Estimation}

Galaxy spectra test subtler physical inference: the PROVABGS labels~\citep{hahn2023provabgs} summarize a galaxy's stellar population and formation history and test whether the representation supports subtler physical inference. The regression targets are: Specific star formation rate (sSFR), mass-weighted age ($t_{\mathrm{age}}$), mass-weighted metallicity ($Z_{\mathrm{MW}}$), and stellar mass ($\log M_*$). OmniSpectra outperforms in most cases (Table~\ref{tab:downstream_combined}), substantially for three of the four targets.
% This split matches the intended division of labor: attention pooling preserves local morphology, wavelength anchoring preserves where it occurs.

\begin{table}[!ht]
    \centering
    \small
    \setlength{\tabcolsep}{1.5pt}
    \begin{tabular*}{\columnwidth}{l@{\extracolsep{\fill}}cccccccc}
    \toprule
         & \multicolumn{2}{c}{sSFR} & \multicolumn{2}{c}{$t_{\mathrm{age}}$} & \multicolumn{2}{c}{$Z_{\mathrm{MW}}$} & \multicolumn{2}{c}{$\log M_*$} \\
    \cmidrule(lr){2-3}\cmidrule(lr){4-5}\cmidrule(lr){6-7}\cmidrule(lr){8-9}
         \textbf{Model} & $R^2$ & $\sigma^\mathrm{MAD}$ & $R^2$ & $\sigma^\mathrm{MAD}$ & $R^2$ & $\sigma^\mathrm{MAD}$ & $R^2$ & $\sigma^\mathrm{MAD}$ \\
    \midrule
        \ourrow \textbf{OmniSpectra} & \textbf{0.72} & \textbf{1.00} & \textbf{0.55} & \textbf{0.57} & \textbf{0.69} & \textbf{0.41} & \textbf{0.89} & \underline{0.20} \\
  AION-1 & 0.64 & 1.13 & 0.47 & 0.62 & 0.60 & 0.49 & \underline{0.88} & \underline{0.20} \\
  AstroCLIP & 0.62 & 1.15 & 0.46 & \underline{0.60} & 0.61 & 0.48 & 0.86 & 0.22 \\
  Shen et al. & 0.65 & 1.11 & 0.49 & 0.67 & 0.63 & 0.46 & 0.86 & 0.21 \\
  Perceiver & 0.63 & 1.18 & 0.49 & 0.67 & 0.60 & 0.48 & 0.84 & 0.23 \\
         \midrule
    GalSpecNet & \underline{0.67} & \underline{1.10} & \underline{0.50} & 0.65 & 0.64 & 0.46 & \underline{0.88} & \textbf{0.18} \\
    Spender & \underline{0.67} & 1.12 & 0.48 & 0.65 & \underline{0.65} & \underline{0.43} & 0.87 & 0.22 \\
    \bottomrule
    \end{tabular*}
    \caption{Galaxy property estimation on DESI PROVABGS ($R^2$\,$\uparrow$, $\sigma^\mathrm{MAD}$\,$\downarrow$).}
    
\label{tab:downstream_combined}
\end{table}

\subsection{Source Classification}
Classification asks whether the tokens also support semantic separation. Distinguishing stars, galaxies, and quasars is complicated by class imbalance and by features moving with redshift. OmniSpectra achieves the best F1 for every SDSS class (Table~\ref{tab:src_class_f1}) and again leads with limited labels (Fig.~\ref{fig:low_data_regime}(b)). Even with as few as $<$500 labeled samples, OmniSpectra shows significant performance. With more labeled data, other models slowly keep closing the gap. 

\begin{table}[!ht]
\centering
\small
\setlength{\tabcolsep}{3.5pt}
\begin{tabular}{lccc}
\toprule
\textbf{Model} & \textbf{Galaxy} & \textbf{Quasar} & \textbf{Star} \\
\midrule
\ourrow \textbf{OmniSpectra} & \textbf{0.99} & \textbf{0.93} & \textbf{0.97} \\
AION-1 & 0.87 & 0.05 & 0.40 \\
AstroCLIP & 0.98 & \underline{0.92} & 0.92 \\
Shen et al. & 0.98 & 0.89 & 0.92 \\
Perceiver & 0.98 & 0.89 & 0.93 \\ \midrule
GalSpecNet & 0.94 & 0.54 & 0.45 \\
Spender & \underline{0.99} & 0.89 & \underline{0.94} \\
\bottomrule
\end{tabular}
\caption{Per-class source classification. F1~($\uparrow$) on SDSS.}
\label{tab:src_class_f1}
\end{table}

% \begin{table}[!ht]
% \centering
% \small
% \setlength{\tabcolsep}{3.5pt}
% \begin{tabular}{lccc}
% \toprule
% \textbf{Model} & \textbf{Galaxy} & \textbf{Quasar} & \textbf{Star} \\
% \midrule
% \ourrow \textbf{OmniSpectra} & \textbf{0.99$_{\pm0.00}$} & \textbf{0.93$_{\pm0.01}$} & \textbf{0.97$_{\pm0.01}$} \\
% AION-1 & 0.87$_{\pm0.01}$ & 0.05$_{\pm0.01}$ & 0.40$_{\pm0.02}$ \\
% AstroCLIP & 0.98$_{\pm0.00}$ & \underline{0.92$_{\pm0.01}$} & 0.92$_{\pm0.02}$ \\
% Shen et al. & 0.98$_{\pm0.00}$ & 0.89$_{\pm0.02}$ & 0.92$_{\pm0.01}$ \\
% Perceiver & 0.98$_{\pm0.00}$ & 0.89$_{\pm0.03}$ & 0.93$_{\pm0.01}$ \\ \midrule
% GalSpecNet & 0.94$_{\pm0.04}$ & 0.54$_{\pm0.29}$ & 0.45$_{\pm0.27}$ \\
% Spender & \underline{0.99$_{\pm0.00}$} & 0.89$_{\pm0.01}$ & \underline{0.94$_{\pm0.01}$} \\
% \bottomrule
% \end{tabular}
% \caption{Per-class source classification. F1~($\uparrow$) on SDSS, mean$_{\pm\mathrm{std}}$ over 5 seeds.}
% \label{tab:src_class_f1}
% \end{table}

\subsection{Redshift Estimation}

\begin{figure*}[!ht]
  \centering
  \begin{subfigure}{0.45\textwidth}
    \includegraphics[width=\textwidth]{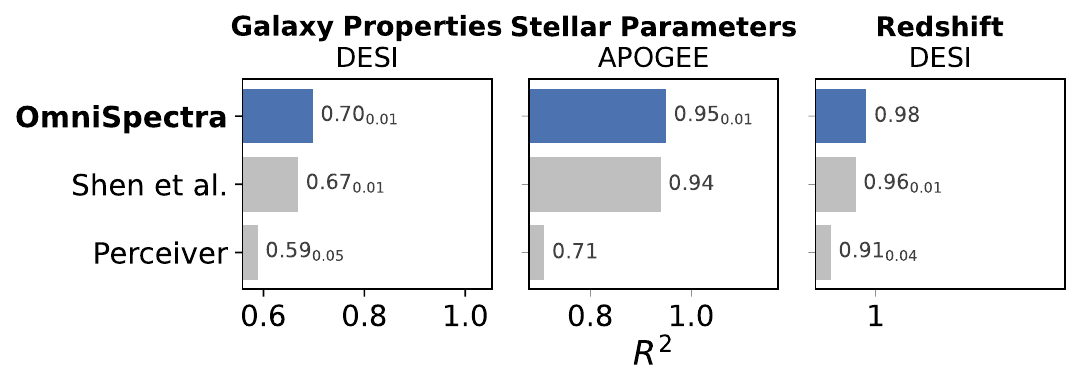}
    \subcaption{Cross-instrument transfer. Average $R^2$ under SDSS-only pretraining with $|\Delta R^2|$ relative to full-corpus.}
    \label{fig:cross_survey_ablation}
  \end{subfigure}
  \hfill
  \begin{subfigure}{0.54\textwidth}
    \includegraphics[width=\textwidth]{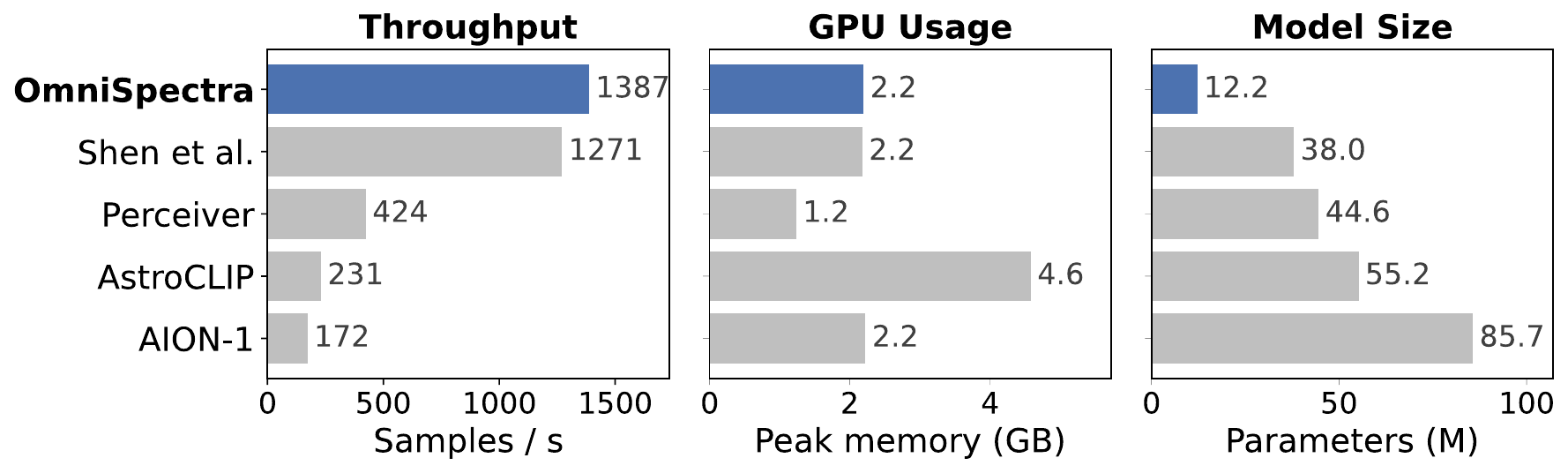}
    \subcaption{Inference efficiency on the DESI dataset. Throughput, peak GPU memory, and active parameters.}
    \label{fig:inference_benchmark}
  \end{subfigure}
  \caption{Transfer robustness (left) and computational efficiency (right).}
  \label{fig:transfer_efficiency}
\end{figure*}

\begin{figure}[!ht]
  \centering
  \includegraphics[width=\linewidth]{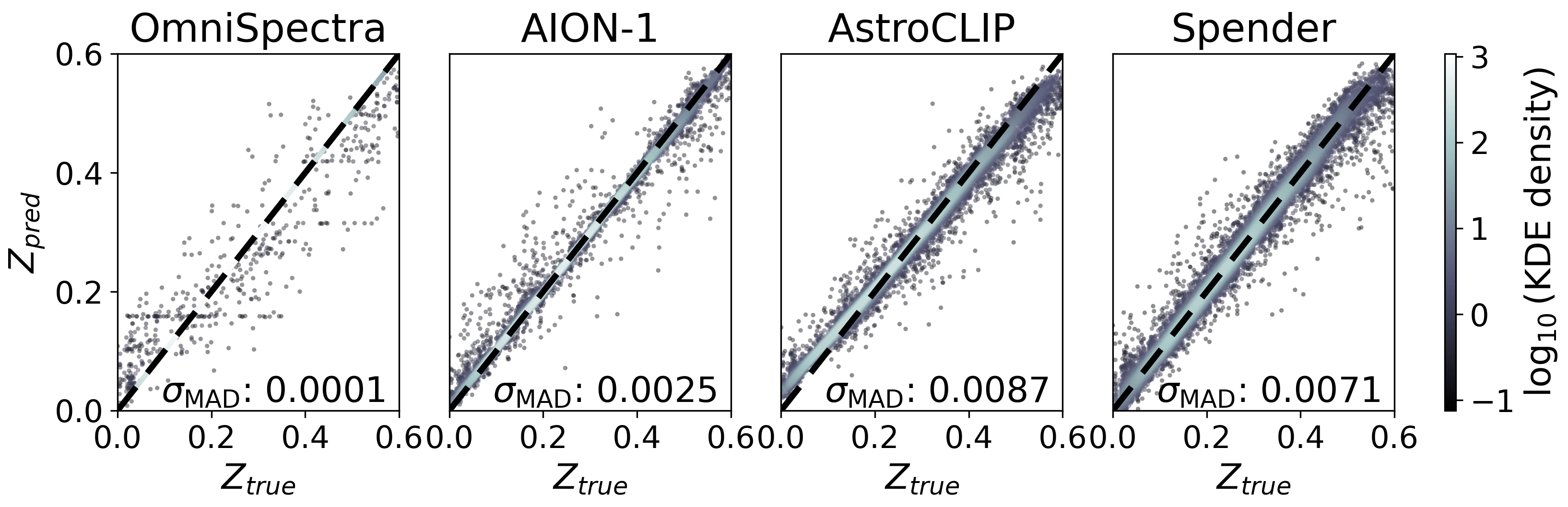}
  \caption{Redshift estimation on DESI PROVABGS.}
  % Top: predicted vs.\ true redshift with $\sigma^\mathrm{MAD}$ inset. Bottom: normalized residual $(z_\mathrm{pred}-z_\mathrm{true})/(1+z_\mathrm{true})$ vs.\ true redshift.
  \label{fig:redshift_comparison}
\end{figure}

Table~\ref{tab:redshift_baselines} and Fig.~\ref{fig:redshift_comparison} report $R^2$, $\sigma^\mathrm{MAD}$, and outlier rate on DESI PROVABGS. OmniSpectra achieves the best $R^2$-score and ties GaSNet-III on $\sigma^\mathrm{MAD}$, at a nearly $5\times$ lower outlier rate. Most models get a good $R^2$-score; $\sigma^\mathrm{MAD}$ is where many of them struggle. Omnispectra achieves that while also keeping the catastrophic outliers low. 
% Our reported redshift is a coarse-to-fine hybrid: OmniSpectra's frozen-probe estimate windows GaSNet-III's template search (Appendix~E of the technical appendix).
% Perceiver lags on all three metrics, confirming that integer-rank positional assumptions do not transfer across grids.

\begin{table}[!ht]
    \centering
    \small
    \begin{tabular}{lccc} \toprule
        \textbf{Model} & $R^2$ ($\uparrow$) & $\sigma^\mathrm{MAD}$ ($\downarrow$) & Outlier\,\% ($\downarrow$) \\ \midrule
        \ourrow \textbf{OmniSpectra} & \textbf{0.99} & \textbf{0.0001} & \textbf{0.06} \\
        AION-1       & \underline{0.99} & \underline{0.01} & 0.07 \\
        AstroCLIP    & 0.98 & 0.01 & 0.06 \\
        Shen et al.  & 0.95 & 0.01 & 0.11 \\
        Perceiver & 0.95 & 0.01 & 0.14 \\ 
        GaSNet-III   & 0.97 & \textbf{0.0001} & 0.28 \\
        \midrule
        Spender      & 0.98 & 0.01 & 0.06 \\
        GalSpecNet   & 0.98 & 0.01 & 0.06 \\
        \bottomrule
    \end{tabular}
    \caption{Redshift estimation on DESI PROVABGS ($Z_\mathrm{HP}$). Top: pretrained; bottom: supervised.}
    \label{tab:redshift_baselines}
\end{table}

\subsection{Ablation Study}
\label{sec:ablation}

Table~\ref{tab:ablation} isolates one component at a time within the same architecture, replacing with its conventional alternative while holding the rest at their proposed setting: mean pooling of linearly projected pixels (tokenizer), a learned sequence-index embedding in place of log-$\lambda$+RoPE (positional code), and a per-patch broadcast cross-attention decoder without Fourier basis or gate (reconstruction); $-$Gate keeps the Fourier decoder but drops LSF-dependent attenuation. \textbf{Base} uses all alternatives, \textbf{Full} all proposed components. Contributions are task-dependent and non-additive: the positional code matters most on APOGEE, consistent with transfer from the optical pretraining range to near-infrared, and the LSF gate on source classification; some partial configurations underperform \textit{Base}, indicating component interactions.

\begin{table}[!ht]
    \centering
    \small
    \setlength{\tabcolsep}{3.2pt}
    \begin{tabular}{lcccccG} \toprule
    \textit{Component} & Base & $-$Tok & $-$PE & $-$Dec & $-$Gate & \textbf{Full} \\
    Tokenizer          & -- & --  & \OK & \OK & \OK & \OK \\
    Positional Code    & -- & \OK & --  & \OK & \OK & \OK \\
    Implicit Decoder   & -- & \OK & \OK & --  & \OK & \OK \\
    LSF Gate           & -- & \OK & \OK & --  & --  & \OK \\
    \midrule
    \multicolumn{7}{l}{\textit{Task Performance}} \\
    Stellar $R^2$   & 0.544 & 0.568 & 0.560 & 0.561 & 0.581 & \textbf{0.599} \\
    APOGEE $R^2$    & 0.797 & 0.929 & 0.759 & 0.923 & 0.929 & \textbf{0.940} \\
    Source F1       & 0.968 & 0.980 & 0.975 & 0.973 & 0.964 & \textbf{0.985} \\
    Galaxy $R^2$    & 0.668 & 0.680 & 0.670 & 0.687 & 0.686 & \textbf{0.713} \\
    Redshift $R^2$  & 0.965 & 0.976 & 0.959 & 0.980 & 0.981 & \textbf{0.991} \\
    \bottomrule
    \end{tabular}
    \caption{Leave-One-Component-Out Ablation. \OK\ enables a component and -- substitutes its standard alternative.}
    \label{tab:ablation}
\end{table}

\subsection{Cross-Survey Generalization}
\label{sec:cross_instrument}

We pretrain the three variable-length models (OmniSpectra, Shen et al., Perceiver) on SDSS alone (3800--9200\,\AA) and evaluate on DESI and APOGEE with no further pretraining. APOGEE (15100--17000\,\AA) lies entirely outside that range, testing transfer to unseen physical coordinates. OmniSpectra nonetheless transfers with little degradation (Fig.~\ref{fig:cross_survey_ablation}): both wavelength-encoding models stay robust, OmniSpectra with the higher average, whereas Perceiver's index-based encoding degrades sharply on redshift.

\subsection{Inference Efficiency}
\label{sec:efficiency}

Native-pixel patching scales encoder sequence length with the observed spectrum rather than a fixed grid. On identical DESI batches (Fig.~\ref{fig:inference_benchmark}, batch size 64), OmniSpectra and Shen et al. process shorter sequences and reach the highest throughput, OmniSpectra with far less memory than per-pixel and fixed-length baselines. For AstroCLIP (370M) and AION-1 (314M), we count only parameters active on spectral input.

\subsection{Limitations}
The positional encoding is defined as a continuous function of log-wavelength and can therefore be evaluated outside the wavelength union of the training instruments without architectural modification. However, its period range is tuned to approximately \(3{,}600\)–\(18{,}000\,\text{\AA}\), and performance at substantially different wavelengths has not been evaluated. The learned representations are currently unimodal; extending them to other modalities, for example through CLIP-style alignment, remains future work.

\section{Conclusion}
\label{sec:conclusion}

OmniSpectra tackles a pervasive alignment challenge in continuous physical signals: coordinate mismatch across heterogeneous instruments. Built on a standard transformer, it replaces integer-indexed embeddings with wavelength-native codes: anchoring position to physical wavelength gives the same feature the same code on any grid, while log-$\lambda$ turns redshift into an additive shift that cancels in RoPE's relative attention. This reaches state-of-the-art results on classification, redshift, and property regression in full- and low-data regimes with only 12.2M parameters. The gains come from encoding structure, not pretraining breadth: physics-grounded inductive biases (log-wavelength translation, hardware resolution gating) offer foundation models a more efficient and robust route to alignment.

% \section{Acknowledgements}
% We acknowledge support from the National Science Foundation under Cooperative Agreement 2421782, the Simons Foundation through award MPS-AI-00010515 and NSF Simons Cosmic AI Seed Grant AWD-006703. We are grateful for the resources and support provided by UVA Research Computing.

\bibliography{bibliography}

\end{document}